\documentclass[
reprint,
superscriptaddress,
amsmath,amssymb,
apl,
]{revtex4-2}

\usepackage[utf8]{inputenc}
\usepackage{graphicx,import}
\graphicspath{{Figures/}}
\usepackage{bm}
\usepackage[colorlinks]{hyperref}
\usepackage[all]{hypcap}
\usepackage{xcolor}
\usepackage{braket}
\usepackage{siunitx}
\usepackage{changes}
\usepackage{lipsum}
\usepackage{nicefrac}
\usepackage{booktabs}
\usepackage{colortbl}
\usepackage{listings}

\definecolor{maroon}{cmyk}{0, 0.87, 0.68, 0.32}
\definecolor{halfgray}{gray}{0.55}
\definecolor{ipython_frame}{RGB}{207, 207, 207}
\definecolor{ipython_bg}{RGB}{247, 247, 247}
\definecolor{ipython_red}{RGB}{186, 33, 33}
\definecolor{ipython_green}{RGB}{0, 128, 0}
\definecolor{ipython_cyan}{RGB}{64, 128, 128}
\definecolor{ipython_purple}{RGB}{170, 34, 255}

\usepackage{listings}
\lstdefinelanguage{iPython}{
    morekeywords={access,and,break,class,continue,def,del,elif,else,except,exec,finally,for,from,global,if,import,in,is,lambda,not,or,pass,print,raise,return,try,while},%
    morekeywords=[2]{abs,all,any,basestring,bin,bool,bytearray,callable,chr,classmethod,cmp,compile,complex,delattr,dict,dir,divmod,enumerate,eval,execfile,file,filter,float,format,frozenset,getattr,globals,hasattr,hash,help,hex,id,input,int,isinstance,issubclass,iter,len,list,locals,long,map,max,memoryview,min,next,object,oct,open,ord,pow,property,range,raw_input,reduce,reload,repr,reversed,round,set,setattr,slice,sorted,staticmethod,str,sum,super,tuple,type,unichr,unicode,vars,xrange,zip,apply,buffer,coerce,intern},%
    sensitive=true,%
    morecomment=[l]\#,%
    morestring=[b]',%
    morestring=[b]",%
    morestring=[s]{'''}{'''},% used for documentation text (mulitiline strings)
    morestring=[s]{"""}{"""},% added by Philipp Matthias Hahn
    morestring=[s]{r'}{'},% `raw' strings
    morestring=[s]{r"}{"},%
    morestring=[s]{r'''}{'''},%
    morestring=[s]{r"""}{"""},%
    morestring=[s]{u'}{'},% unicode strings
    morestring=[s]{u"}{"},%
    morestring=[s]{u'''}{'''},%
    morestring=[s]{u"""}{"""},%
    literate=
    {á}{{\'a}}1 {é}{{\'e}}1 {í}{{\'i}}1 {ó}{{\'o}}1 {ú}{{\'u}}1
    {Á}{{\'A}}1 {É}{{\'E}}1 {Í}{{\'I}}1 {Ó}{{\'O}}1 {Ú}{{\'U}}1
    {à}{{\`a}}1 {è}{{\`e}}1 {ì}{{\`i}}1 {ò}{{\`o}}1 {ù}{{\`u}}1
    {À}{{\`A}}1 {È}{{\'E}}1 {Ì}{{\`I}}1 {Ò}{{\`O}}1 {Ù}{{\`U}}1
    {ä}{{\"a}}1 {ë}{{\"e}}1 {ï}{{\"i}}1 {ö}{{\"o}}1 {ü}{{\"u}}1
    {Ä}{{\"A}}1 {Ë}{{\"E}}1 {Ï}{{\"I}}1 {Ö}{{\"O}}1 {Ü}{{\"U}}1
    {â}{{\^a}}1 {ê}{{\^e}}1 {î}{{\^i}}1 {ô}{{\^o}}1 {û}{{\^u}}1
    {Â}{{\^A}}1 {Ê}{{\^E}}1 {Î}{{\^I}}1 {Ô}{{\^O}}1 {Û}{{\^U}}1
    {œ}{{\oe}}1 {Œ}{{\OE}}1 {æ}{{\ae}}1 {Æ}{{\AE}}1 {ß}{{\ss}}1
    {ç}{{\c c}}1 {Ç}{{\c C}}1 {ø}{{\o}}1 {å}{{\r a}}1 {Å}{{\r A}}1
    {€}{{\EUR}}1 {£}{{\pounds}}1
    {^}{{{\color{ipython_purple}\^{}}}}1
    {=}{{{\color{ipython_purple}=}}}1
    {+}{{{\color{ipython_purple}+}}}1
    {*}{{{\color{ipython_purple}$^\ast$}}}1
    {/}{{{\color{ipython_purple}/}}}1
    {+=}{{{+=}}}1
    {-=}{{{-=}}}1
    {*=}{{{$^\ast$=}}}1
    {/=}{{{/=}}}1,
    literate=
    *{-}{{{\color{ipython_purple}-}}}1
     {?}{{{\color{ipython_purple}?}}}1,
    identifierstyle=\color{black}\ttfamily,
    commentstyle=\color{ipython_cyan}\ttfamily,
    stringstyle=\color{ipython_red}\ttfamily,
    keepspaces=true,
    showspaces=false,
    showstringspaces=false,
    numberstyle=\tiny\color{halfgray},
    basicstyle=\scriptsize,
    keywordstyle=\color{ipython_green}\ttfamily,
}

\usepackage{mdframed}
\mdfdefinestyle{PseudocodeFrame}{%
    linecolor=ipython_frame,
    roundcorner=10pt,
    innertopmargin=0pt,
    innerbottommargin=0pt,
    innerrightmargin=5pt,
    innerleftmargin=5pt,
    backgroundcolor=ipython_bg
}

\definecolor{bluegray}{RGB}{40,180,160}
\definecolor{navygray}{RGB}{110,140,170}
\definecolor{meadowgreen}{RGB}{0,128,0}
\definecolor{coolbrown}{RGB} {165,42,42}
\definecolor{lightred}{RGB} {242,190,190}
\definecolor{lightgreen}{RGB} {191,226,191}

\DeclareSIUnit{\sq}{\Box}

\newcommand{\secref}[1]{\hyperref[#1]{{Sec.~\ref{#1}}}}
\newcommand{\chapref}[1]{\hyperref[#1]{{Chap.~\ref{#1}}}}
\newcommand{\suppref}[1]{\hyperref[#1]{{App.~\ref{#1}}}}
\newcommand{\figref}[1]{\hyperref[#1]{{Fig.~\ref*{#1}}}}
\newcommand{\Figref}[1]{\hyperref[#1]{{Figure~\ref*{#1}}}}
\newcommand{\figrefadd}[2]{\hyperref[#1]{{Fig.~\ref*{#1}#2}}}
\newcommand{\tabref}[1]{\hyperref[#1]{Table~\ref*{#1}}}
\renewcommand{\eqref}[1]{\hyperref[#1]{{Eq.~\ref*{#1}}}}
\newcommand{\Eqref}[1]{\hyperref[#1]{{Equation~\ref*{#1}}}}

\let\itext\i
\renewcommand{\i}{\mathrm{i}}
\newcommand{\e}{\mathrm{e}}
\newcommand{\Qi}{Q_\mathrm{i}}
\newcommand{\Qc}{Q_\mathrm{c}}
\newcommand{\Ql}{Q_\mathrm{l}}
\renewcommand{\wr}{\omega_\text{r}}
\newcommand{\fr}{f_\text{r}}

\newcommand{\revise}[1]{{#1}}
\newcommand{\revisex}[1]{{}}

\hypersetup{
citecolor={navygray}, 
linkcolor={navygray},
urlcolor={navygray},
}

\makeatletter
\newcommand*{\balancecolsandclearpage}{%
  \close@column@grid
  \cleardoublepage
  \twocolumngrid
}
\makeatother

\begin{document}
\title{Fano Interference in Microwave Resonator Measurements}

\author{D.~Rieger}
\email{dennis.rieger@kit.edu}
\affiliation{PHI,~Karlsruhe~Institute~of~Technology,~76131~Karlsruhe,~Germany}

\author{S.~Günzler}
\thanks{First two authors contributed equally.}
\affiliation{PHI,~Karlsruhe~Institute~of~Technology,~76131~Karlsruhe,~Germany}
\affiliation{IQMT,~Karlsruhe~Institute~of~Technology,~76344~Eggenstein-Leopoldshafen,~Germany} 

\author{M.~Spiecker}
\affiliation{PHI,~Karlsruhe~Institute~of~Technology,~76131~Karlsruhe,~Germany}
\affiliation{IQMT,~Karlsruhe~Institute~of~Technology,~76344~Eggenstein-Leopoldshafen,~Germany}

\author{A.~Nambisan}
\affiliation{PHI,~Karlsruhe~Institute~of~Technology,~76131~Karlsruhe,~Germany}
\affiliation{IQMT,~Karlsruhe~Institute~of~Technology,~76344~Eggenstein-Leopoldshafen,~Germany}

\author{W.~Wernsdorfer}
\affiliation{PHI,~Karlsruhe~Institute~of~Technology,~76131~Karlsruhe,~Germany}
\affiliation{IQMT,~Karlsruhe~Institute~of~Technology,~76344~Eggenstein-Leopoldshafen,~Germany}

\author{I.~M.~Pop}
\email{ioan.pop@kit.edu}
\affiliation{PHI,~Karlsruhe~Institute~of~Technology,~76131~Karlsruhe,~Germany}
\affiliation{IQMT,~Karlsruhe~Institute~of~Technology,~76344~Eggenstein-Leopoldshafen,~Germany}

\date{\today}

\begin{abstract}
Resonator measurements are a simple but powerful tool to characterize a material's microwave response. The losses of a resonant mode are quantified by its internal quality factor $\Qi$, which can be extracted from the scattering coefficient in a microwave reflection or transmission measurement. Here we show that a systematic error on $\Qi$ arises from Fano interference of the signal with a background path. Limited knowledge of the interfering paths in a given setup translates into a range of uncertainty for $\Qi$, which increases with the coupling coefficient. We experimentally illustrate the relevance of Fano interference in typical microwave resonator measurements and the associated pitfalls encountered in extracting $\Qi$. On the other hand, we also show how to characterize and utilize the Fano interference to eliminate the systematic error.
\end{abstract}

\maketitle
\section{Introduction}
Understanding and mitigating dissipation in superconducting quantum hardware is an ongoing endeavor spanning from material science to microwave engineering~\cite{Krantz2019, Siddiqi2021}. The characterization of various loss channels~--~such as dielectric, inductive or quasiparticle-induced -- is experimentally accessible by using microwave resonators. Compared to qubits, resonators are easier to design, fabricate, measure and analyze. For this reason, microwave resonator loss measurements are routinely used when conducting material studies \cite{Gao2007, Palacios-Laloy2008, Barends2008, Vissers2010, Paik2010, Weber2011, Reagor2013, Minev2013, Dupre2017, Gruenhaupt2018, Shearrow2018, Verjauw2021,Verjauw2021, Gao2022}, addressing particular loss channels \cite{Barends2010, Barends2011, Geerlings2012, Vissers2012, Nsanzineza2014, Patel2017, Brehm2017, Henriques2019} or comparing different fabrication recipes, shielding, etc. \cite{Barends2011, Bruno2015, Calusine2018, deGraaf2018, Melville2020, Altoe2022}. Commonly, the loss budget is determined by measuring the dependence of losses on design, bias or drive parameters (e.g. the dielectric participation ratio \cite{Wenner2011, Wang2015, Gruenhaupt2018}, electric/magnetic field \cite{Sarabi2016, Borisov2020}, readout power \cite{Khalil2010}, etc.).

While the resonance frequency and total linewidth of a mode can be extracted from its frequency response, the task of separating the intrinsic losses of the mode from the coupling to the external measurement apparatus is more delicate. These contributions to the linewidth are quantified by the internal ($\Qi$) and coupling (or external) quality factor ($\Qc$), respectively, and they are encoded in the complex-plane geometry of the resonator response. But what if the measurement is distorted systematically by imperfections in the microwave setup? One frequent example in the literature is the occurrence of asymmetric lineshapes, which is attributed to impedance mismatches at the input and output ports~\cite{Khalil2011,Probst2015}. Consequently, the extracted internal losses depend on the model used to account for the experimental imperfections.

In this article, we argue that a systematic source of uncertainty for microwave resonator scattering experiments arises from Fano interference~\cite{Fano_original} of the resonant signal with background paths, for example due to finite circulator or sample holder port-to-port isolation. As a consequence, the scattering data becomes intertwined with the amplitude and phase of the Fano interference, which are generally unknown. Remarkably, this holds true even for apparently undistorted measurements with symmetric lineshapes. An unambiguous extraction of the coupling coefficient $\Qi/\Qc$ is therefore impossible and, instead, we suggest a procedure to extract a range of uncertainty based on an upper bound for the interference amplitude. This range increases with the coupling strength, which we illustrate by measuring a set of 9 resonators in both the undercoupled ($\Qi<\Qc$) and overcoupled ($\Qi>\Qc$) regime. Moreover, we show that by characterizing the Fano interference we can lift the uncertainty in the $\Qi$ measurement.

This manuscript is structured as follows: In the next section we illustrate the effects of Fano interference on resonator scattering measurements. In \secref{sec:transform}, we calculate the scattering coefficient transformation due to Fano interference. \secref{sec:error_procedure} proposes a method to solve the inverse problem of calculating internal losses from experimental data. Finally, in \secref{sec:measurements} we apply the method to a set of resonators measured in different coupling regimes.

\section{Origin and Symptoms of Fano Interference}
\label{sec:origin}
\begin{figure*}[t!]
\centering
\includegraphics{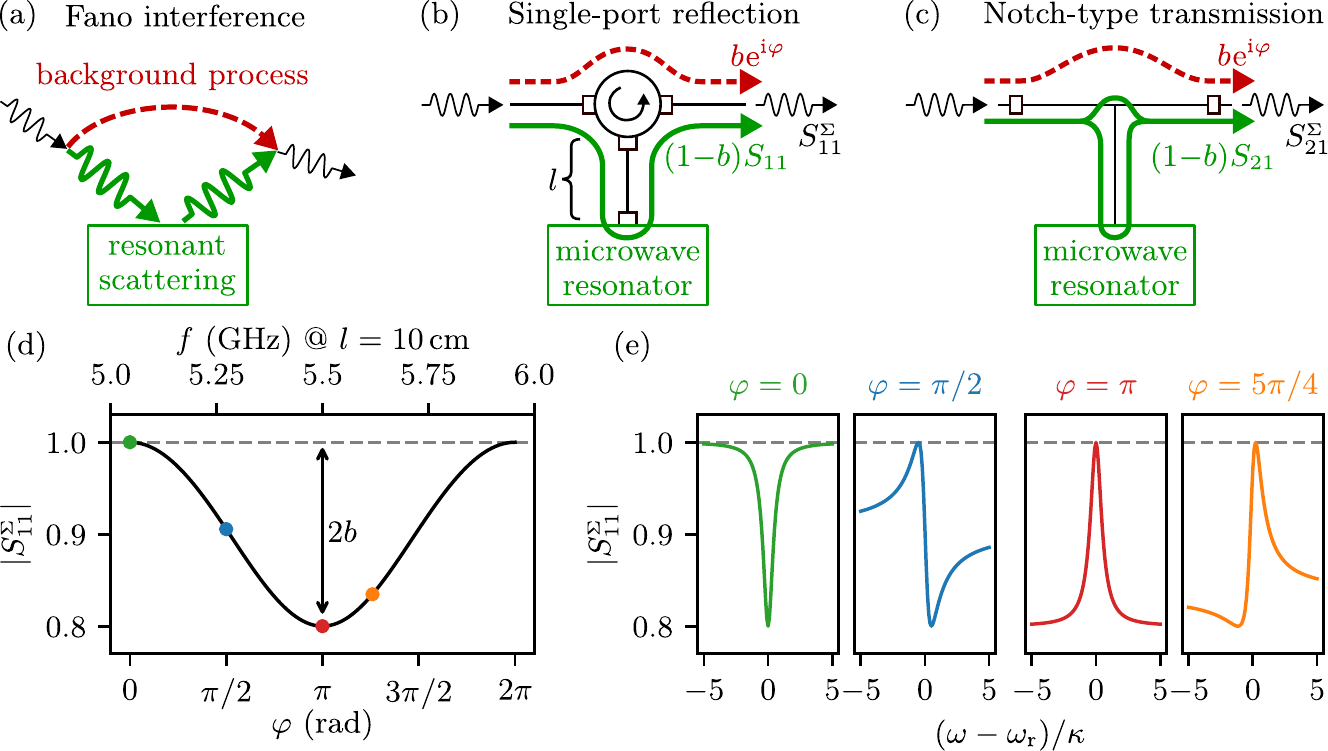}
\caption{\textbf{Origin and symptoms of Fano interference in microwave resonator measurements.} \textbf{(a)}~Schematic depiction of Fano interference in wave scattering experiments: A background path (red dashed arrow) interferes with the signal scattered by the measured resonant system (green arrows). \textbf{(b)}~\revise{Fano interference in }\revisex{While present in all microwave setups (see main text), here we focus the discussion on }single-port reflection measurements where the finite circulator isolation constitutes a dominant background path: the leakage signal (red arrow), defined by a relative amplitude $b$ and phase $\varphi$, interferes with the signal reflected by the sample $(1-b)S_{11}$ (green arrow).
\revise{Fano interference in hanger-type transmission measurements where two-port crosstalk gives rise to the leakage signal $b \e^{\i\varphi}$ (red arrow).}
\revise{\textbf{(d)}}~Background interference pattern: Away from resonance the reflection coefficient equals unity ($S_{11}=1$, dashed gray line); however, the measured baseline $|S_{11}^{_\Sigma}|$ varies by $\pm b$ depending on $\varphi$. The $\varphi \in [0,2\pi)$ interval can be mapped to a corresponding (periodic) frequency range for a given optical path length $l$ between the sample and the circulator (cf.~(b)). The top axis shows an example frequency interval for $l=\SI{10}{\centi\meter}$ (cf.~\eqref{eq:backgroundFreq}). \revise{\textbf{(e)}}~Fano lineshapes: Resonator responses $|S_{11}^{_\Sigma}|$ can be asymmetric or even exhibit peaks depending on $\varphi$ (indicated by the colored markers in \revise{(d)}). Note that in this calculation we assumed lossless modes ($|S_{11}|=1$, dashed gray line).
}
\label{fig:origin}
\end{figure*}

Fano interference is generally observed in wave scattering experiments~\cite{Fano_original,rau2004perspectives} and it occurs whenever a resonantly scattered signal interferes with background paths (\figrefadd{fig:origin}{a}). Such paths are inherently present in microwave resonator measurement setups due to port-to-port leakage, impedance mismatches or other device imperfections. In the following, we focus the discussion on single-port reflection measurements, in which the finite circulator isolation constitutes a dominant background path (\figrefadd{fig:origin}{b}). Using complex phasor language, the background path with amplitude~$b$ and phase~$\varphi$ (relative to the resonant signal) interferes with the reduced scattered signal $(1-b)S_{11}$ to give the measured signal
\begin{equation}
    S_{11}^{_\Sigma} = (1-b)S_{11} + b\!\:\e^{\i\varphi}\,.
    \label{eq:fanoPhasor}
\end{equation}
We note that \revise{in \eqref{eq:fanoPhasor}, except for the case of perfectly constructive interference, part of the signal is reflected into the input port, to preserve energy conservation. While \eqref{eq:fanoPhasor} only considers a single background phasor $b\!\:\e^{\i\varphi}$, additional paths can lead to a functional dependence of the measured signal on $S_{11}$ with more degrees of freedom.} \revisex{(e.g. reflections at the coaxial cable connection to the circulator or sample holder). }Moreover, we would like to emphasize that background paths leading to Fano interference are also present in transmission measurements \revise{(\figrefadd{fig:origin}{c})}, which we \revisex{briefly }discuss in \suppref{sec:supp:transmission}.

In general, $S_{11}$, $b$ and $\varphi$ in \eqref{eq:fanoPhasor} are frequency dependent. Before analyzing resonant modes ($S_{11}(\omega)$) we first consider the effect of the interference on the microwave line background far away from resonances ($S_{11}=1$) for constant $b$. In this case, the relative interference phase $\varphi$ is given by the optical path length between the interfering signals, and the amplitude of the measured signal oscillates by $\pm b$ as a function of $\varphi$ (\figrefadd{fig:origin}{d}). Assuming that the optical path length is given by a coaxial microwave cable of length $l$ between sample and circulator, the phase interval $\varphi\in[0,2\pi)$ can be mapped to a frequency interval,
\begin{equation}
    \frac{\varphi}{2\pi}=2 l\frac{ \Delta f}{c_\text{eff}}\approx\frac{l}{\SI{10}{\centi\meter}}\frac{\Delta f}{\SI{1}{\giga\hertz}}\,, \label{eq:backgroundFreq}
\end{equation}
where $c_\text{eff}\approx 0.7 c$ is the speed of light in the coaxial cable ($\epsilon_r\approx 2.1$). A typical length $l =\SI{10}{\centi\metre}$ corresponds to a \SI{1}{\giga\hertz} period for $\varphi$. While these oscillations carry information about the Fano phasors, in practice this information is often hidden by other contributions to the microwave line background, such as standing wave patterns from impedance mismatches \revise{(cf.~\suppref{sec:supp:microwavebackground})}.

The scattering coefficient of the sample $S_{11}$ changes around the resonant frequency $\wr$ within a span on the order of the linewidth $\kappa$ (${\sim\si{\mega\hertz}}$), much smaller than the $\varphi(f)$ period (therefore $\varphi\approx\text{const.}$). In the limit of a lossless response, the phase of the single-port reflection coefficient $S_{11}(\omega)$ is swept by $2\pi$,
\begin{subequations}
\begin{align}
    |S_{11}|&=1\label{eq:losslessResponseAmplitude}\\
    \arg S_{11}(\omega) &= -2\arctan\left(\frac{2(\omega-\wr)}{\kappa}\right)\label{eq:losslessResponsePhase}\,,
\end{align}
\end{subequations}
which takes $S_{11}^{_\Sigma}$ through one complete interference pattern. As shown in \figrefadd{fig:origin}{e}, despite using lossless modes in the calculation, for which we expect a flat amplitude response (dashed line in \figrefadd{fig:origin}{e}), the resulting amplitude lineshapes show dips, peaks and different degrees of asymmetry depending on the relative background phase $\varphi$. These apparent loss and gain curves illustrate the systematic distortions of the resonator response due to Fano interference, which make it difficult to extract an accurate value for $\Qi$. Moreover, measuring a symmetric lineshape is no guarantee for the absence of Fano interference (cf. $\varphi=0,\pi$ in \figrefadd{fig:origin}{e}). \revise{Note, that for hanger-type transmission measurements (cf.~\figrefadd{fig:origin}{c}) asymmetric amplitude lineshapes are commonly interpreted as a result of impedance mismatches \cite{Khalil2011, Deng2013, Probst2015}. In contrast, the asymmetry in \figrefadd{fig:origin}{e} emerges solely from Fano interference. Impedance mismatches in the signal path $S_{21}$ (green arrow in \figrefadd{fig:origin}{c}) could further distort the amplitude lineshapes (cf.~\suppref{sec:supp:transmission} for detailed discussion).} In order to quantify the uncertainty \revise{resulting from Fano interference} for resonator loss measurements, in the next section we consider resonators in reflection with finite intrinsic losses ($|S_{11}|<1$).

\section{Transformation of Scattering Data due to Fano Interference}
\label{sec:transform}
\begin{figure*}[t!]
\centering
\includegraphics[width=\textwidth]{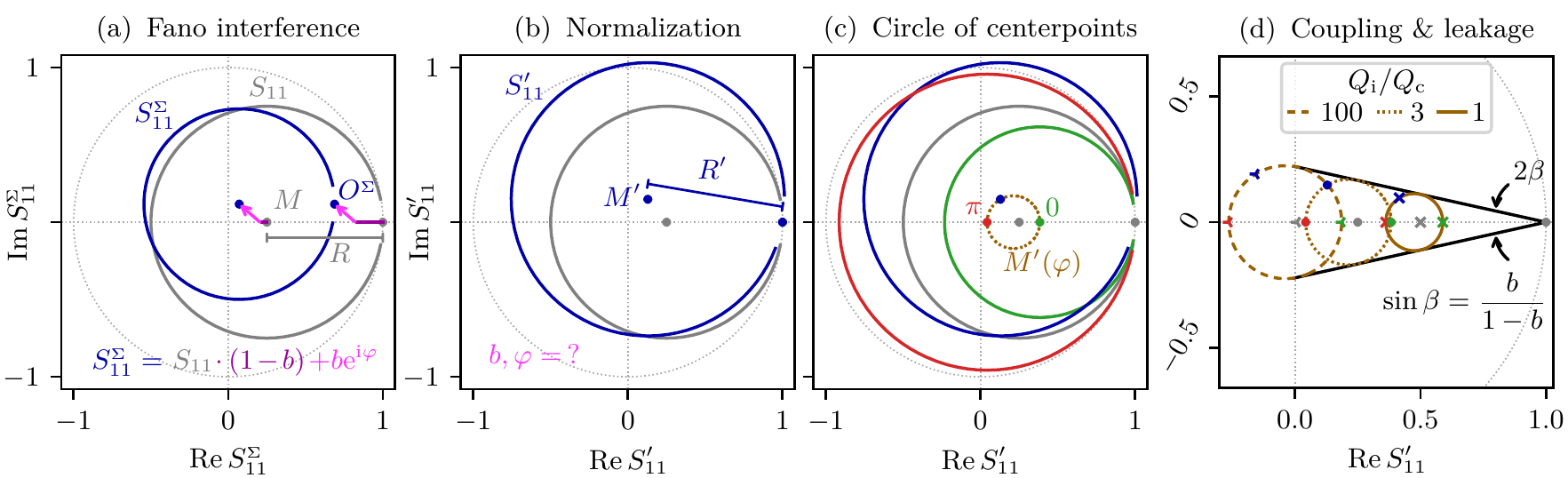}
\caption{\textbf{Transformation of scattering data in the complex plane due to Fano interference.} In all panels, the axes through the origin and the unit circle are shown for reference in dotted, light gray. \textbf{(a)}~The intrinsic scattering data $S_{11}$ (gray circle with center point $M$ and radius $R$) is scaled by $1-b$ and shifted by $b\e^{\i\varphi}$ (magenta arrows), where $b$ and $\varphi$ are the amplitude and phase of the Fano leakage phasor, respectively (cf.~\figrefadd{fig:origin}{b\revise{,c}} and~\eqref{eq:fanoPhasor}). \textbf{(b)}~Experimentally, the leakage phasor is -- in general -- inaccessible. Therefore, the data is normalized to the measured off-resonant point (labeled $O^{_\Sigma}$ in panel (a)) shown by the blue circle. The normalized scattering data $S_{11}'$ is fully described by its center point $M'$. Notably, Fano interference gives rise to a tilted circle with altered radius $R'$ and center point $M'$ (compare blue and gray circles). \textbf{(c)}~For different phases $\varphi$, the transformed center points $M'$ describe a circle $M'(\varphi)$ (\revise{dotted, }ocher) corresponding to different tilts and radii of $S_{11}'$. Note that constructive (green) and destructive (red) Fano interference lead to untilted $S_{11}'$ circles (i.e. symmetric amplitude lineshape, cf. \figrefadd{fig:origin}{e}). \textbf{(d)}~The coupling coefficient $\Qi/\Qc$ scales the horizontal position of the $M'(\varphi)$-circle with respect to the off-resonant point. For fixed leakage amplitude $b$, the $M'(\varphi)$-circles are bounded by two lines (in black) at an angle $\sin\beta=\tilde{b}$ with the real axis (cf.~\eqref{eq:FanoConeAngle}). As a consequence, an ambiguity arises for inferring the coupling coefficient from a measurement: Any measured center point $M'$ can be attributed to one of two $M'(\varphi)$-circles intersecting at $M'$, as illustrated by the blue disc marker and the $M'(\varphi)$ circles for $\Qi/\Qc=3$ (\revise{dotted}) and $\Qi/\Qc=100$ (dashed). Note that in all panels $b=0.18$, corresponding to a circulator isolation of $\SI{15}{\deci\bel}$.
}
\label{fig:fanoTransform}
\vspace{0.3cm}
\end{figure*}
\begin{figure}[t!]
\centering
\includegraphics[width=\columnwidth]{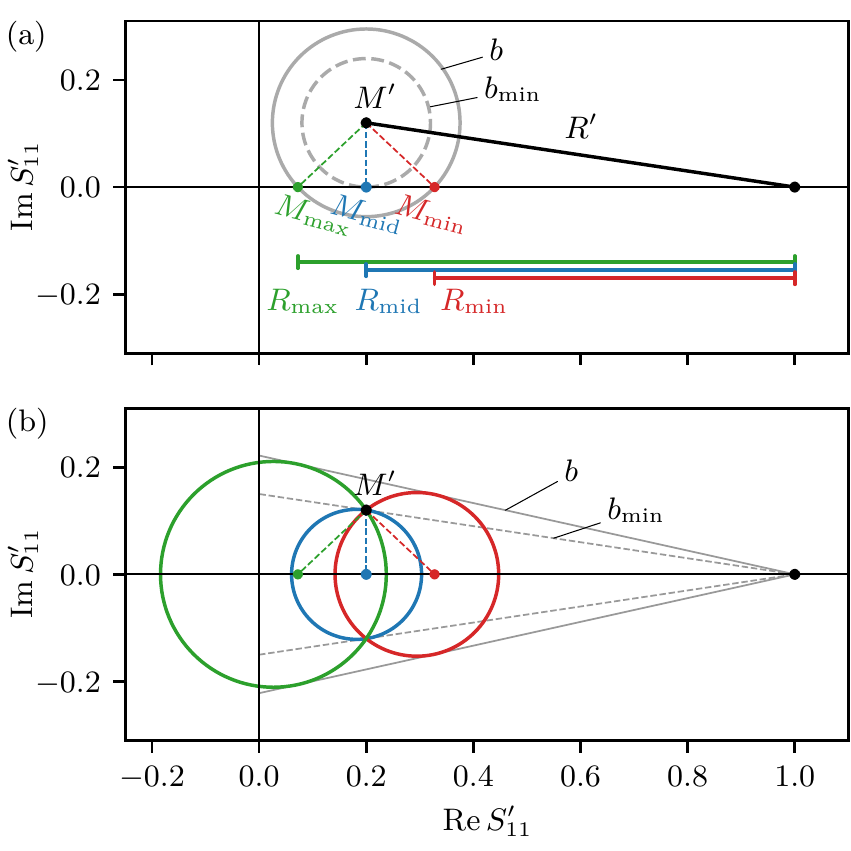}
\caption{\textbf{Inferring the range for the coupling coefficient.}
\textbf{(a)}~Center points $M$ consistent with a measured center point $M'$ (cf. \figrefadd{fig:fanoTransform}{b}) are given by intersections of circles (grey) with radius $|R'| b$ around $M'$ and the real axis (cf.~\eqref{eq:inverseCircle}). Assuming an upper bound $b$ for the leakage amplitude (see main text), two solutions $M_\text{min}$ (red), $M_\text{max}$ (green) exist and the corresponding radii $R_\text{min}$, $R_\text{max}$ bound the range for the coupling coefficient~$\Qi/\Qc$ (cf.~\eqref{eq:RadiusQ}). The smallest circle (dashed grey) around $M'$ touching the real axis in a single point $M_\text{mid}$ (blue) corresponds to the minimum leakage amplitude $b_\text{min}$ consistent with $M'$. Importantly, the associated $(\Qi/\Qc)_\text{mid}$ is the median of the coupling coefficients within the range $(\Qi/\Qc)_\text{min}$ and $(\Qi/\Qc)_\text{max}$ (because $M_\text{min}$ and $M_\text{max}$ are symmetric around $M_\text{mid}$). Therefore, $(\Qi/\Qc)_\text{mid}$ serves as a reference value independent of the assumed $b$. \textbf{(b)}~Circles of center points $M'(\varphi)$ around the solutions $M$ derived in panel (a). Evidently, all circles intersect at $M'$. The circles around $M_\text{min}$ (red) and $M_\text{max}$ (green) are the smallest and largest circles, respectively, which contain $M'$ and are bounded by the lines for $b$ (solid gray). The circle around $M_\text{mid}$ (blue) contains $M'$ with the minimum leakage amplitude $b_\text{min}$ (dashed gray).
}
\label{fig:errorprocedures}
\end{figure}
\subsection{Extracting Internal Losses from Scattering Data}
In the vicinity of a resonance, the reflection coefficient describes a circle in the complex plane,
\begin{align}
    &S_{11}(\omega)=1-\frac{2\Ql/\Qc}{1 + 2\i\Ql(\omega-\wr)/\wr}\,,\label{eq:S11circle}\\
    &\text{with}\quad \frac{1}{\Ql} =  \frac{1}{\Qc} + \frac{1}{\Qi} \,,\label{eq:QlQcQi}
\end{align}
where $\wr$ is the resonance frequency and $\Qi$, $\Qc$ and $\Ql$ are the internal, coupling and loaded quality factor, respectively. Extracting the internal losses from measured data consists of two steps: First, the frequency response yields the resonance frequency $\wr$ and the total linewidth
\begin{equation}
    \kappa=\frac{\wr}{\Ql}\label{eq:kappa}\,,
\end{equation}
fixing the loaded quality factor. Second, the radius $R$ of the circle in the complex plane,
\begin{equation}
    R=\frac{\Ql}{\Qc}=\frac{\Qi/\Qc}{\Qi/\Qc+1}\,, \label{eq:RadiusQ}
\end{equation}
determines the coupling coefficient $\Qi/\Qc$ and, therefore, the contributions of coupling and internal losses to the loaded quality factor in \eqref{eq:QlQcQi}.

In practice, the amplitude of the scattering data is scaled due to attenuation/amplification in the input/output lines and the circle is rotated around the origin by a phase linked to the propagation delay through the measurement setup. These transformations can be accounted for by normalizing the data to the off-resonant response ($\omega-\wr\gg\kappa$) in the complex plane such that the off-resonant point is located at $(1,0)$. As a consequence, the response circle is entirely defined by its radius or, equivalently, its center point. Unfortunately, Fano interference introduces a translation of the data in the complex plane, which can not be corrected by normalization. 

\subsection{Fano Transform}
In \figref{fig:fanoTransform} we show how Fano interference transforms the scattering data in the complex plane. Compared to the scattered signal $S_{11}$ in the absence of Fano (gray circle in \figrefadd{fig:fanoTransform}{a}), the measured signal $S_{11}^\Sigma$ (blue dashed circle) is scaled by $(1-b)$ and translated by the leakage phasor $b\e^{\i\varphi}$ (magenta arrow) according to \eqref{eq:fanoPhasor}. This reasoning assumes knowledge of both interfering phasors (i.e. $S_{11}$, $b$, $\varphi$), which in practice is usually not the case. From a measurement perspective we access $S_{11}^{_\Sigma}$, which only carries information about the sum of the phasors. Since we do not know the off-resonant response of $S_{11}$, we have to normalize the data to the measured off-resonant response $O^{_\Sigma}$, which yields
\begin{align}
    S_{11}'=\frac{(1-b)S_{11}+b\e^{\i\varphi}}{1-b+b\e^{\i\varphi}}=\frac{S_{11}+\tilde{b}\e^{\i\varphi}}{1+\tilde{b}\e^{\i\varphi}}\label{eq:fanoTransform}\,,
\end{align}
where we introduced
\begin{equation}
    \tilde{b}=\frac{b}{1-b}\label{eq:btilde}
\end{equation}
as the amplitude of the Fano background phasor relative to the maximum amplitude $1-b$ at the sample port.

Note that \eqref{eq:fanoTransform} is the normalized version of \eqref{eq:fanoPhasor} and implements a Möbius transform of $S_{11}$, which implies that $S_{11}'$ remains a circle and conserves its linewidth $\kappa$. The normalized scattering data (\figrefadd{fig:fanoTransform}{b}, blue circle) is tilted with respect to the off-resonant point $(1, 0)$, i.e. its center point $M'$ is displaced from the real axis. For different interference phases $\varphi$, the center point describes a circle $M'(\varphi)$ (ocher in \figrefadd{fig:fanoTransform}{c}, see~\suppref{sec:supp:circleofcenterpoints} for derivation) corresponding to different tilts and radii of the normalized measured data. From the construction of the $S_{11}'$ circles in \figrefadd{fig:fanoTransform}{c} it is clear that in general the measured lineshapes are asymmetric (cf.~\figrefadd{fig:origin}{e}) with the exception of the two cases of constructive ($\varphi=0$) or destructive ($\varphi=\pi$) Fano interference.

Finally, we are now ready to discuss the dependence of $S_{11}'$ on internal losses and coupling. In the absence of Fano interference, the center points $M$ of the normalized $S_{11}$ response are scaled on the real axis as a function of the coupling coefficient, as illustrated by the gray markers in \figrefadd{fig:fanoTransform}{d}. Similarly, in the presence of Fano interference, the circles of center points $M'(\varphi)$ are scaled (dilated) with respect to the off-resonant point, as a function of the coupling coefficient. For a leakage amplitude $b$, the circles are bounded by two lines at angles $\pm\beta$, with
\begin{equation}
    \sin\beta = \frac{b}{1-b}=\tilde{b}\label{eq:FanoConeAngle}\,,
\end{equation}
as illustrated in \figrefadd{fig:fanoTransform}{d} for $\Qi/\Qc$ of $1$, $3$ and $100$. Evidently, the lack of information on the leakage phasor results in multiple $M'(\varphi)$ circles and corresponding coupling coefficients consistent with a measured center point $M'$. This is illustrated in \figrefadd{fig:fanoTransform}{d} by the $M'(\varphi)$ circles for $\Qi/\Qc=3$ and $100$ intersecting at the blue disc marker. As a consequence of this ambiguity, the true internal losses cannot be inferred from measured data without further information or assumptions. During the editing of this manuscript we became aware of a note~\cite{Steele_Fano_lineshapes} by Gary Steele, which argues along similar lines and reaches the same conclusion.

\vspace{1cm}
\section{Systematic Uncertainty Range for the Coupling Coefficient}
\label{sec:error_procedure}
\begin{figure}[t!]
\centering
\includegraphics[width=\columnwidth]{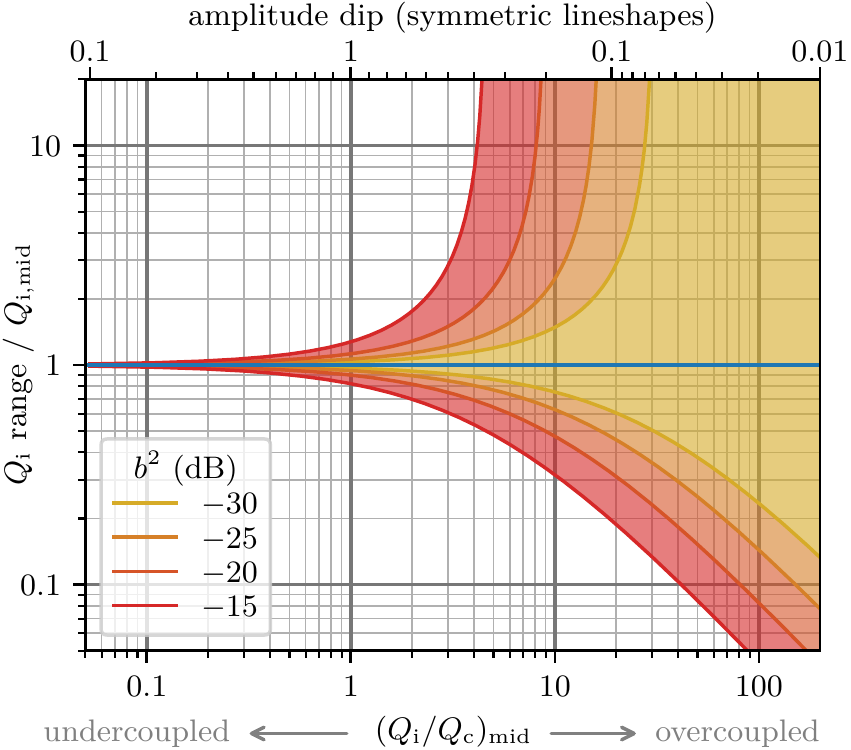}
\caption{\textbf{Maximum inferred $\Qi$ uncertainty range vs. coupling regime.}
The bands show the range of systematic uncertainty for $\Qi$ relative to the median $Q_\text{i,mid}$ (blue line) based on the procedure described in \figref{fig:errorprocedures}. The color encodes different upper bounds for the leakage amplitude $b$ for selected circulator isolations listed in the legend. For critical coupling $\left(\Qi / \Qc \right)_\text{mid} = 1$, the relative uncertainty is on the order of $b$ and it increases significantly towards stronger coupling. Crucially, the probability distribution inside the band for each coupling coefficient is non-Gaussian; the $Q_\text{i,mid}$ is the median of this distribution. The bands are calculated for $M'$ on the real axis (symmetric resonance lineshapes), which lead to maximum uncertainty.  For convenience, the top axis shows the measured amplitude dip size $1-|S^\prime_{11}(\wr)|$.}
\label{fig:errorbands}
\end{figure}

Direct access to the Fano leakage phasor is challenging because the signal paths of interest (cf.~\figrefadd{fig:origin}{b\revise{,c}}) are part of the extended microwave setup which includes frequency dependent components, impedance mismatches and standing waves. While it is possible to identify the background phasor via an in-situ experimental calibration (cf.~\secref{sec:measurements} or Ref.~\cite{Wang2021Jun}), it requires dedicated hardware in the setup. Therefore, in order to quantify internal losses in the presence of Fano interference, we need to make assumptions on the background amplitude~$b$ or phase~$\varphi$. Inferring $\varphi$ based on the measured oscillations of the background is generally unreliable \revise{(cf.~\suppref{sec:supp:microwavebackground})}. On the other hand, estimating an upper bound for the maximum interference amplitude is more feasible, for example by using the nominal leakage stated in the data sheets of commercial circulators. For typical \SIrange{4}{12}{\giga\hertz} single junction circulators the upper bound can be taken to be $b^2=-\SI{15}{\dB}$, i.e. $b=0.18$ \cite{LNF2022}.

\vspace{0.3cm}
In the following paragraphs, we derive step by step how the assumption of a maximum leakage amplitude bounds the systematic uncertainty for the inferred internal losses $\Qi$. We start by inverting the Fano transform \eqref{eq:fanoTransform},
\vspace{0.2cm}
\begin{align}
    M &= M' + (M'-1)\tilde{b}\e^{\i\varphi}\nonumber\\
    &= M' - R' \tilde{b}\e^{\i\varphi}\nonumber \\
    &= M' - |R'| \tilde{b}\e^{\i\varphi+\mathrm{arg}R'}\,, \label{eq:inverseCircle}
\end{align}
where we identify $1-M'$ as the vector $R'$ from $M'$ to the off-resonant point (cf.~\figrefadd{fig:errorprocedures}{a}). \Eqref{eq:inverseCircle} describes a circle around $M'$ with radius $|R'|\tilde{b}$ (solid gray). In the absence of Fano interference the center point $M$ is located on the real axis (cf.~\figrefadd{fig:fanoTransform}{a}), therefore, the solutions $M$ consistent with the measured $M'$ are given by the intersections of the circle~\eqref{eq:inverseCircle} with this axis: $M_\text{min}$ (green), $M_\text{max}$ (red). 
For decreasing background amplitude $b$, i.e. decreasing radius of the gray circle (\eqref{eq:inverseCircle}), the inferred center points move closer until for a leakage amplitude $b=b_\text{min}$ the gray dashed circle touches the real axis in a single point $M_\text{mid}$. This $M_\text{mid}$ is the projection of $M'$ on the real axis and the solution explaining the measured $M'$ with minimum leakage amplitude (cf.~\figrefadd{fig:errorprocedures}{b}). Due to symmetry, for uniformly sampled interference phase, $M_\text{mid}$ is the median of the distribution of possible center points between $M_\text{min}$ and $M_\text{max}$. Note that the range between $M_\text{min}$ and $M_\text{max}$ is maximized for $M'$ measured on the real axis (the green and red circles become tangent in \figrefadd{fig:errorprocedures}{b}). For a given leakage $b$, this leads to the largest uncertainty for the inferred internal losses $\Qi$.

The $S_{11}$ circle radii for each of the points $M_\text{min}$, $M_\text{mid}$, $M_\text{max}$ are given by
\begin{align}
    R_\text{min,max} &= \mathrm{Re}\:\!R' \mp  \sqrt{(|R'| \tilde{b})^2 - \mathrm{Im}^2\:\!R'} \label{eq:inverseTransformR} \nonumber\\
    &=R_\text{mid} \mp |R'| \sqrt{\tilde{b}^2 - \tilde{b}_\text{min}^2} \, ,
\end{align} 
and they can be converted to the corresponding coupling coefficients using \eqref{eq:RadiusQ}. Using this information and the fact that $\Ql$ is not affected by Fano interference, we calculate the resulting relative uncertainty for $\Qi$ as a function of the coupling regime. We consider the case in which $M'$ is measured on the real axis (maximum uncertainty). In \figref{fig:errorbands} we plot the relative uncertainty bands $\Qi/Q_\text{i,mid}$ as a function of $(\Qi/\Qc)_\text{mid}$ for four values of $b$, as indicated by the different shades of orange. The bands widen with increasing coupling coefficient and Fano amplitude. Note that as a consequence of the non-linear relation \eqref{eq:RadiusQ} between $R$ and the coupling coefficient, the $\Qi$ uncertainty range is asymmetric with respect to $Q_\text{i,mid}$ and the distribution of solutions inside the range is neither uniform nor Gaussian. Nevertheless, $Q_\text{i,mid}$ remains the median.

In practice, \figref{fig:errorbands} serves as a valuable reference chart to read off the expected uncertainty for a measurement with a particular coupling coefficient. For convenience we also show the corresponding amplitude dip size $1-|S_{11}'(\wr)|$  as the top axis. The relative uncertainty is on the order of $b$ (i.e. few percent) for critical coupling and it increases substantially towards stronger coupling. We would like to highlight the relevance of $Q_\text{i,mid}$ as a reference value, as it indicates the median of the uncertainty distribution independent of the leakage amplitude~$b$. Consequently, in the analysis of experimental results with our method, internal losses are always quantified by $\Qi$ ranges from $Q_\text{i,min}$ to $Q_\text{i,max}$ and the median value at $Q_\text{i,mid}$. We note that previous cicle fit algorithms for transmission such as Ref.~\cite{Khalil2011,Geerlings2012,Probst2015, Deng2013} result in single $\Qi$ estimates \revisex{close to }\revise{equivalent to} $Q_\text{i,mid}$ (cf.~\suppref{sec:supp:transmission}). In \suppref{sec:supp:pseudocode} we illustrate by means of pseudocode how to integrate our method to extract the $\Qi$ uncertainty in a circle fit routine. Moreover, we provide an example implementation and analysis script in the repository \cite{Rieger2023Mar}.

\section{Application to Measurements}
\label{sec:measurements}
We illustrate our method to extract the $\Qi$ uncertainty in the presence of Fano interference by measuring granular aluminum (grAl) resonators in both the overcoupled and undercoupled regime, because they correspond to different $\Qi$ uncertainty intervals (cf.~\figref{fig:errorbands}). Using lift-off e-beam lithography we patterned nine grAl stripline resonators on a c-plane sapphire chip. The film thickness is \SI{20}{\nano\meter} and its sheet resistance is \SI{1.5}{\kilo\ohm/\sq}. The resonators are \SI{4}{\micro\meter} wide and, by varying their length between \SI{400}{\micro\meter} and \SI{950}{\micro\meter}, their resonant frequencies span a range of \SIrange{5}{11.3}{\giga\hertz}. Details on sample parameters and the sample holder are listed in \suppref{sec:supp:samples}. In the following we discuss three types of $\Qi$ measurements which impose different constraints on the Fano amplitude $b$ and phase $\varphi$. 

\begin{figure*}[t!]
\centering
\includegraphics[width=\textwidth]{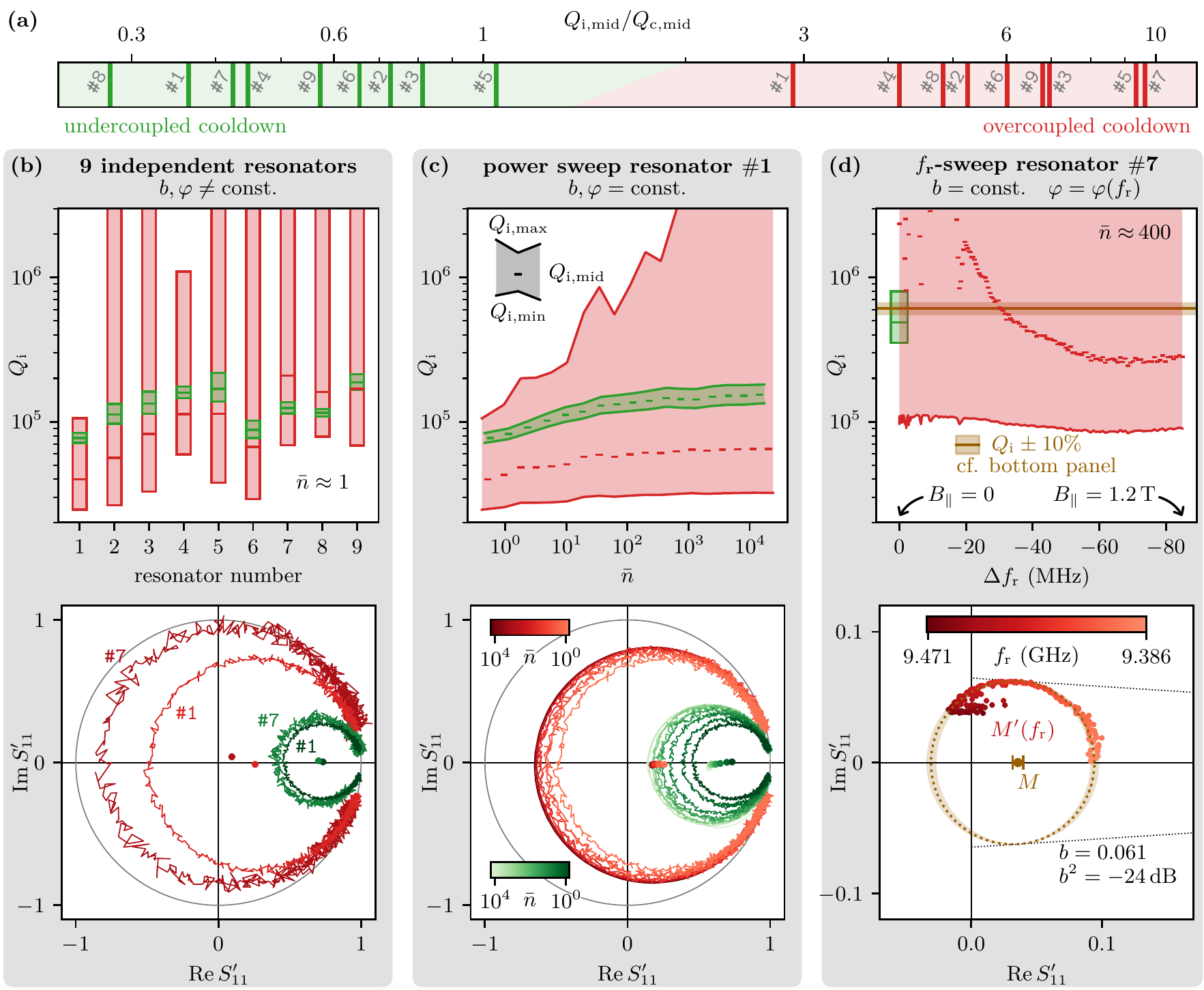}
\caption{\textbf{\revise{Typical internal loss measurements: interpreting and minimizing the uncertainty range.}}
\revise{\textbf{(a)}~Coupling coefficient $Q_\text{i,mid}/Q_\text{c,mid}$ for the 9 resonators measured in (b) -- (d). For the same samples, in subsequent cooldowns we decrease the $\Qi$ uncertainty range by reducing the coupling from overcoupled (red) to undercoupled (green).}
The columns \textbf{(\revise{b})--(\revise{d})} showcase \revise{three} measurement scenarios with different constraints on the Fano amplitude $b$ and phase $\varphi$\revise{, and for each type of measurement we compare data for the two coupling regimes (red, green)}. The top panels show the $\Qi$ ranges %\revisex{(extracted following the procedure in \figref{fig:errorprocedures}, using $b^2=-\SI{15}{\deci\bel}$~\cite{LNF2022})}
as intervals extending from $Q_\text{i,min}$ to $Q_\text{i,max}$ and including a horizontal line at the median value $Q_\text{i,mid}$ \revise{(extracted following the procedure in \figref{fig:errorprocedures}, using $b^2=-\SI{15}{\deci\bel}$~\cite{LNF2022})}. In the bottom panels we present corresponding scattering data in the complex plane. \revisex{For the same samples, in subsequent cooldowns we decrease the $\Qi$ uncertainty range by reducing the coupling from overcoupled ($\Qc\sim10^4$, red) to undercoupled ($\Qc\sim10^5$, green).} \revise{\textbf{(b)}}~Measurements of 9 independent samples, each with unknown $b$, $\varphi$. Notice that for all resonators the narrower uncertainty ranges of the undercoupled measurements are within the uncertainty ranges for the overcoupled regime. \revise{\textbf{(c)}}~A power sweep is a prime example for a measurement with the restriction $b,\varphi\approx\text{const.}$, since the frequency change is much smaller than the frequency period of the Fano interference (cf.~\figrefadd{fig:origin}{d}). \revise{\textbf{(d)}}~Sweeping the resonator frequency over an interval comparable to the frequency period of the Fano interference can lead to deceptive systematic trends in the inferred $\Qi$ values in the overcoupled regime, due to evolution of the interference phase $\varphi(f)$ (cf.~\figrefadd{fig:origin}{d}). Here we change $f_\text{r}$ by applying an in-plane magnetic field and the measured center points $M'$ trace out a circle $M'(f_\text{r})$ in the complex plane (bottom panel). Remarkably, a fit to this trajectory (dotted circle in ocher) enables the extraction of both the leakage amplitude $b=0.065$ and the true $\Qi$, as indicated by the center point $M$ and the horizontal line (ocher) in the upper panel. The ocher shaded area indicates the $\Qi\pm 10\%$ interval in the top panel and the corresponding $M'(f_\text{r})$ circles in the bottom panel, consistent with the error bar for the position of point $M$.
}
\label{fig:examples}
\vspace{2cm}
\end{figure*}

In \figrefadd{fig:examples}{a,b} we show results measured on all nine resonators at similar powers corresponding to a circulating photon number $\bar{n}\approx1$. In consecutive cooldowns we decouple the same samples \revise{(\figrefadd{fig:examples}{a})} from $\Qc\sim10^4$ (overcoupled, red color scheme) to $\Qc\sim10^5$ (undercoupled, green color scheme) by increasing the distance between the coupling pin and the resonators in our cylindrical waveguide sample holder (cf.~\suppref{sec:supp:samples}). As expected from \figref{fig:errorbands}, in the top panel of \figrefadd{fig:examples}{b} we confirm that the $\Qi$ uncertainty is reduced as the coupling is reduced. The fact that all green ranges are fully contained within the corresponding red ones is consistent with the assumption that there is no significant change of $\Qi$ between the successive cooldowns. Note that if $R_\text{max}$ given by \eqref{eq:inverseTransformR} is larger than 1, this corresponds to an undefined upper bound for the $\Qi$ uncertainty (lossless resonator is consistent with the data). This is the case for the overcoupled resonators 3, 5, 6, 7 and 9 in \figrefadd{fig:examples}{b}.

Importantly, comparing the internal losses of samples with overlapping uncertainty ranges (overcoupled data) might not be conclusive, because each of the nine resonators is subjected to a different leakage phasor $b\e^{\i\varphi}$ at its resonant frequency. For example, resonators 1 and 7 show a deceptive factor of $5$ difference in $Q_\text{i,mid}$ in the overcoupled regime (red), while the undercoupled data (green) reveals only a factor of $1.5$ change. This point is also illustrated by the scattering data in the complex plane (bottom panel in \figrefadd{fig:examples}{b}). Moreover, resonator 7 has higher $Q_\text{i,mid}$ than resonator 5 in the overcoupled regime, while the decoupled values are inverted. We note that these limitations for interpreting overcoupled data apply both when comparing different samples in the same setup (as shown here) and the same sample in different setups (because the values of $b,\varphi$ are generally setup dependent).

Varying the drive power (\figrefadd{fig:examples}{c}) is representative for the class of measurements with the restriction ${b,\varphi\approx\text{const.}}$, since the resonator frequency shift ($\lesssim\si{\mega\hertz}$ Kerr shift \cite{Maleeva2018}) is typically two orders of magnitude smaller than the period of the Fano background (cf.~\figrefadd{fig:origin}{d}). In this case, trends of $\Qi$ are reflected in trends of the inferred $\Qi$ range (i.e. all of $Q_\text{i,min}$, $Q_\text{i,mid}$ and $Q_\text{i,max}$). Since a trend in $\Qi$ changes the coupling coefficient ($\Qc=\text{const.}$), the uncertainty intervals change asymmetrically (cf.~\figref{fig:errorbands}) and the trends of $Q_\text{i,min}$ and $Q_\text{i,max}$ indicate the weakest and strongest possible trends, respectively, as illustrated by the undercoupled data (green).

In measurements sweeping the resonator frequency $\fr$ on the scale of the Fano period we need to take into account the frequency dependence of both Fano amplitude $b(f)$ and phase $\varphi(f)$. Here, we sweep $\fr$ by applying magnetic field up to \SI{1.2}{\tesla} in-plane with the grAl film of the resonator (cf.~\cite{Borisov2020}). The resonance frequency changes by \SI{85}{\mega\hertz}, which is comparable to the expected Fano background period of $\sim\SI{250}{\mega\hertz}$ for the \SI{40}{\centi\meter} path between the circulator and the sample (cf.~\figrefadd{fig:origin}{b} and \eqref{eq:backgroundFreq}). As illustrated in \figrefadd{fig:examples}{d} (top panel), the extracted $Q_\text{i,mid}$ in the overcoupled regime shows a systematic trend over one order of magnitude. However, as we will show below, this trend is deceptive, and in fact, $\Qi$ remains constant.

By plotting the center points $M'$ of the response in the complex plane (\figrefadd{fig:examples}{d} bottom panel) we find that, with the exception of a few points at low field values, the points $M'(f_\text{r})$ trace out a circle (dotted, ocher). This is consistent with a constant leakage amplitude and a fixed center point $M$, indicating a constant $\Qi$ in this frequency range (cf.~\suppref{sec:supp:otherfield} for other cases). We extract $\Qi=5.7\cdot10^5$, shown as a horizontal ocher line in the upper panel of \figrefadd{fig:examples}{d}. Note that this value is within the measured $\Qi$ interval in the undercoupled regime (green). The $M'(f_\text{r})$ points which deviate from the constant $\Qi$ circle are caused by two phenomena: (i) in the vicinity of $\fr=\SI{9.47}{\giga\hertz}$ the $\Qi$ value fluctuates in time, similarly to Refs.~\cite{Klimov2018,Schloer2019,Winkel2020,Rieger2022}, and (ii) at $B_\parallel=\SI{340}{\milli\tesla}$ the resonator frequency $\fr=\SI{9.451}{\giga\hertz}$ matches the Zeeman splitting of $g=2$ spins in the resonator environment, as previously reported in Refs~\cite{Samkharadze2016, Kroll2019, Borisov2020}.
\balancecolsandclearpage

From the radius of $M'(f_\text{r})$ (cf.~\figrefadd{fig:fanoTransform}{d}) we extract a leakage amplitude of $b=0.061$, i.e. $b^2=\SI{-24}{\deci\bel}$, which matches the isolation reported in the data sheet of the circulator for this frequency range \cite{LNF2022}. In summary, measuring a significant part of the $M'(f_\text{r})$ circle allows us to identify the Fano phasor and therefore removes the corresponding $\Qi$ uncertainty intervals.

\section{Conclusion}
\label{sec:conclusion}
We have shown that limited knowledge about the background path of Fano interference introduces a systematic uncertainty for the extraction of internal losses from resonator scattering measurements. We propose to quantify this uncertainty by assuming an upper bound for the Fano leakage amplitude, which determines a range $(Q_\text{i,min},Q_\text{i,max})$ of possible $\Qi$  values and their median $Q_\text{i,mid}$. While only a few percent in the critically coupled regime, the relative uncertainty range rapidly increases for overcoupled samples. Counter-intuitively, symmetric resonant lineshapes do not guarantee an absence of Fano interference and, in fact, they correspond to the largest uncertainty range. Moreover, we illustrated that Fano interference can lead to amplified, weakened or even deceptive trends for $\Qi$ in measurements such as power or frequency sweeps.

On a positive note, we can gain access to the Fano phasor by tuning in-situ the relative phase between the signal and background path. Here, we illustrate this method by using in-plane magnetic field up to \SI{1.2}{\tesla} to shift the resonator over a significant frequency interval compared to the Fano period. In doing so, we reduce the $\Qi$ uncertainty to $\pm\SI{10}{\percent}$, despite operating deep in the overcoupled regime ($\Qi/\Qc\approx30$). Our approach can be further optimized with dedicated hardware in the cryogenic setup, for example by using a tunable phase-shifter.

\section*{Data and Code Availability}
\revise{The raw data, analysis scripts and circle fit code used in this study are publicly available on Zenodo at \url{https://doi.org/10.5281/zenodo.7767046}.}\revisex{ All relevant data are available from the corresponding author upon reasonable request.}

\section*{Acknowledgements}
We are grateful to P. Winkel and F. Valenti for reading an early version of the manuscript. We acknowledge technical support from A. Bacher, A. Eberhardt, M.K. Gamer, J.K. Hohmann, A. Lukashenko and L. Radtke. Funding is provided by the German Ministry of Education and Research (BMBF) within the project GEQCOS (FKZ: 13N15683). A.N. acknowledges financing from the Baden-Württemberg Stiftung within the project QT-10 (QEDHiNet). D.R., S.G., and W.W. acknowledge support from the European Research Council advanced grant MoQuOS (No.~741276). Facilities use was supported by the Karlsruhe Nano Micro Facility (KNMFi) and KIT Nanostructure Service Laboratory (NSL). We acknowledge qKit for providing a convenient measurement software framework.

\bibliography{references}

\balancecolsandclearpage

\onecolumngrid
\section*{Appendices}
\vspace{0.4cm}
\twocolumngrid
\appendix
\section{Analysis for Transmission Measurements}
\label{sec:supp:transmission}
\revise{
Similar to single-port reflection measurements, two-port hanger-type transmission measurements are routinely used to extract internal losses of microwave resonators. Analogous to reflection, the off-resonant response of hanger-type measurements ideally provides information about the baseline response (i.e. the value of $S_{21}=1$) and is used to normalize the data. In experiments, however, asymmetric amplitude lineshapes (or equivalently tilted $S_{21}$ circles in the complex plane) are ubiquitous and imply the presence of measurement imperfections. Currently, they are entirely attributed to impedance mismatches of the transmission lines connecting to the hanger resonator \cite{Khalil2011,Probst2015, Deng2013}. By modeling equivalent circuits including impedance mismatches in series to the resonator circuit, Refs. \cite{Khalil2011, Deng2013} conclude that the aforementioned imperfections only result in a complex loading of the coupling quality factor and do not influence the extraction of $\Qi$. In other words, the centerpoint of the normalized $S_{21}$ circle is only shifted along the imaginary axis, which can be corrected by projecting it to the real axis (denoted "diameter correction method" in \cite{Khalil2011}). The resulting radius $R$ of the $S_{21}$ circle and extracted internal quality factor $\Qi$ coincide with $R_\text{mid}$ and $Q_\text{i,mid}$, respectively, in this manuscript.

While we agree that this implementation of impedance mismatches can be present in hanger measurements, we argue that these models are incomplete: they do not consider additional signal paths in parallel to the resonator, for example leakage from the input to the output port through the ground plane or direct coupling between the bonding pads on the chip. We can identify these kinds of two-port crosstalk as background paths leading to Fano interference with the signal path (cf.~\figrefadd{fig:origin}{c}). Moreover, it is difficult to engineer these contributions to below a level of \SI{-30}{\deci\bel} for typical chip layouts, comparable to the circulator isolation in reflection measurements. For this reason, even in the absence of the impedance mismatches discussed in Ref.~\cite{Khalil2011, Deng2013}, Fano interference leads to a systematic uncertainty on the extraction of internal losses in hanger-type transmission measurements. In the following, we discuss the consequence of a single leakage path in parallel to an ideal hanger-type transmission response.}

In the vicinity of a resonance, the transmission coefficient reads
\begin{align}
    &S_{21}(\omega)=1-\frac{\Ql/\Qc}{1 + 2\i\Ql(\omega-\wr)/\wr}\,,\label{eq:S21circle}\,
\end{align}
which is identical to the single-port reflection coefficient \eqref{eq:S11circle} besides a factor of 2 in the numerator. The radius of $S_{21}$ in the complex plane is connected to the quality factors by
\begin{equation}
    R=\frac{\Ql}{2\Qc}=\frac{1}{2}\frac{\Qi/\Qc}{\Qi/\Qc+1}\,, \label{eq:S21RadiusQ}
\end{equation}
similarly to \eqref{eq:RadiusQ}. Importantly, \eqref{eq:S21circle} and \eqref{eq:S21RadiusQ} only rescale the relation between radius and quality factors by a factor of 2 compared to the main text discussion. Since Fano interference for fixed leakage amplitude $b$ leads to a constant relative error on the radius (\figrefadd{fig:fanoTransform}{d}), the main text results can also be used for the transmission case. Concretely, one can use \eqref{eq:inverseTransformR} to calculate the uncertainty range for the $S_{21}$ radii and \figref{fig:errorbands} does show the relative uncertainty range for $\Qi$ depending on the coupling coefficient. The only difference is that \eqref{eq:S21RadiusQ} (instead of \eqref{eq:RadiusQ}) must be used to convert between radius and coupling coefficient and, consequently, the range $\Qi/\Qc\in(\revise{0.1}, 100)$ (bottom axis of \figref{fig:errorbands}) maps to an amplitude dip size of $(\revise{0.91}, 0.01)$ in notch-type transmission.

\revise{
We emphasize that, in general, the contributions of Fano interference and impedance mismatches in hanger-type transmission measurements are difficult to disentangle since both imperfections can translate the scattering data in the complex plane. Concretely, the circle of centerpoints and a shift along the imaginary axis become intertwined, which adds an additional degree of freedom that is usually unknown in experiment. For this reason, hanger-type transmission measurements should be treated with a very conservative estimate of the systematic error, i.e. a large upper bound for the leakage amplitude $b$.
}

\revise{
\section{Microwave Baseline}
\label{sec:supp:microwavebackground}
\begin{figure}[tb!]
\centering
\includegraphics[width=\columnwidth]{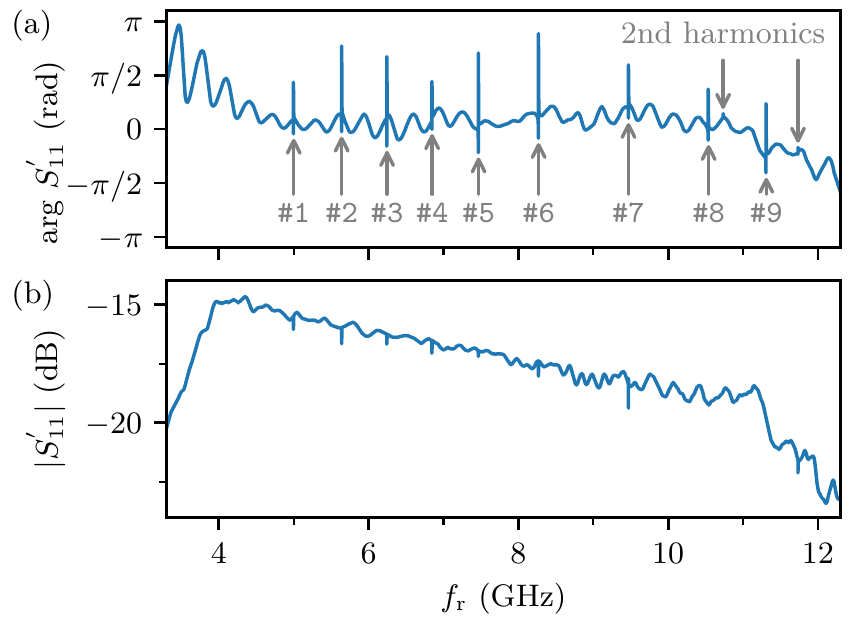}
\caption{\revise{\textbf{Microwave Baseline.} \textbf{(a)} Phase and \textbf{(b)} amplitude of the reflection coefficient $S_{11}'$ covering the $\SIrange{4}{12}{\giga\hertz}$ frequency range of the circulator. The gray arrows mark the frequencies of the 9 resonators discussed in \figref{fig:examples} (overcoupled cooldown).}
}
\label{fig:supp:microwavebackground}
\end{figure}

In \figref{fig:supp:microwavebackground} we show the measured microwave baseline $S_{11}'$, covering the $\SIrange{4}{12}{\giga\hertz}$ range of the circulator used for the measurements in \figref{fig:examples} in the main text. Generally, the amplitude $|S_{11}'|$ decreases with higher frequency (due to an increasing attenuation of the microwave cables) and outside of the frequency range of the microwave components. We emphasize that the variation of $|S_{11}'|$ is below \SI{1}{\deci\bel} within the circulator frequency range~\cite{LNF2022}. In principle, the oscillations of $S_{11}'$ bear resemblance to the expected Fano background interference pattern (\figrefadd{fig:origin}{d}), especially at frequencies at which their oscillation period matches in phase and amplitude (e.g. in the range $\SIrange{9.5}{10.5}{\giga\hertz}$). However, this pattern is superimposed with and generally inseparable from additional contributions like standing wave patterns or frequency features of other microwave components. Therefore, assigning the measured oscillations to Fano interference and to a respective leakage amplitude and phase is unreliable.
}

\section{Pseudocode for Circle Fit Including Systematic Uncertainty}
\label{sec:supp:pseudocode}
The procedure to extract the $\Qi$ uncertainty range (cf.~\secref{sec:error_procedure}) can be implemented as an extension of standard circle fit routines, which we illustrate with the following pseudocode:
\begin{mdframed}[style=PseudocodeFrame]
\begin{lstlisting}[language=iPython, mathescape=true, escapeinside={(*}{*)}]
"""
inputs:
- ${\color{maroon}f}$: scattering data frequencies
- ${\color{maroon}S_{11}^{_\Sigma}}$: measured scattering data at frequencies ${\color{maroon}f}$
- ${\color{maroon}b}$: upper bound for Fano background amplitude
    
outputs:
- ${\color{maroon}f_\text{r}}$: resonant frequency
- ${\color{maroon}\Ql}$:$\,\,$loaded quality factor
- ${\color{maroon}Q_\text{i,min},Q_\text{i,mid},Q_\text{i,max}}$: internal quality factor
                    uncertainty range & median
- ${\color{maroon}Q_\text{c,min},Q_\text{c,mid}, Q_\text{c,max}}$:$\,\,$coupling quality factor
                    uncertainty range & median
"""

# Account for cable delay, scale and rotate the
# circle such that off-resonant point is at ${\color{ipython_cyan}(1, 0)}$
$S_{11}'$ = normalize_circle($f$, $S_{11}^{_\Sigma}$)

# Fit circle center point (complex valued)
$M'$ = fit_normalized_circle($S_{11}'$)
$R'\,\,$ = $1 - M'$ # cf. (*\figref{fig:errorprocedures}*)

# Shift circle to origin and use (*\eqref{eq:losslessResponsePhase}*) to extract
# resonant frequency & total linewidth.
$\fr$, $\kappa$ = fit_frequency_response($f$, $S_{11}'-M'$)
$\Ql$ = $\fr/\kappa$

# Calculate ${\color{ipython_cyan}R_\text{min}, R_\text{mid}, R_\text{max}}$ based on (*\eqref{eq:inverseTransformR}*)
$\tilde{b}$ = $b/(1-b)$ # cf. (*\eqref{eq:btilde}*)
$R_\text{mid}$ = $\mathrm{Re}(R')$
$R_\text{err}$ = $\sqrt{(|R'|\tilde{b})^2 - \mathrm{Im}^2(R')}$
$R_\text{min}$ = $R_\text{mid}- R_\text{err}$
$R_\text{max}$ = $R_\text{mid} + R_\text{err}$
    
# Convert radius to quality factors
# based on (*\eqref{eq:QlQcQi}*) and (*\eqref{eq:RadiusQ}*)
for index in ["min", "mid" "max"]:
    $Q_\text{i,index}$ = $\Ql / (1 - R_\text{index})$
    $Q_\text{c,index}$ = $\Ql / R_\text{index}$
    # Handle unphysical radii
    if $R_\text{index} > 1$:
        $Q_\text{i,index}$ = $\infty$
        $Q_\text{c,index}$ = $\Ql$
\end{lstlisting}
\end{mdframed}

\section{Sample Holder and Sample Parameters}
\label{sec:supp:samples}
\begin{figure}[b!]
\centering
\includegraphics[width=\columnwidth]{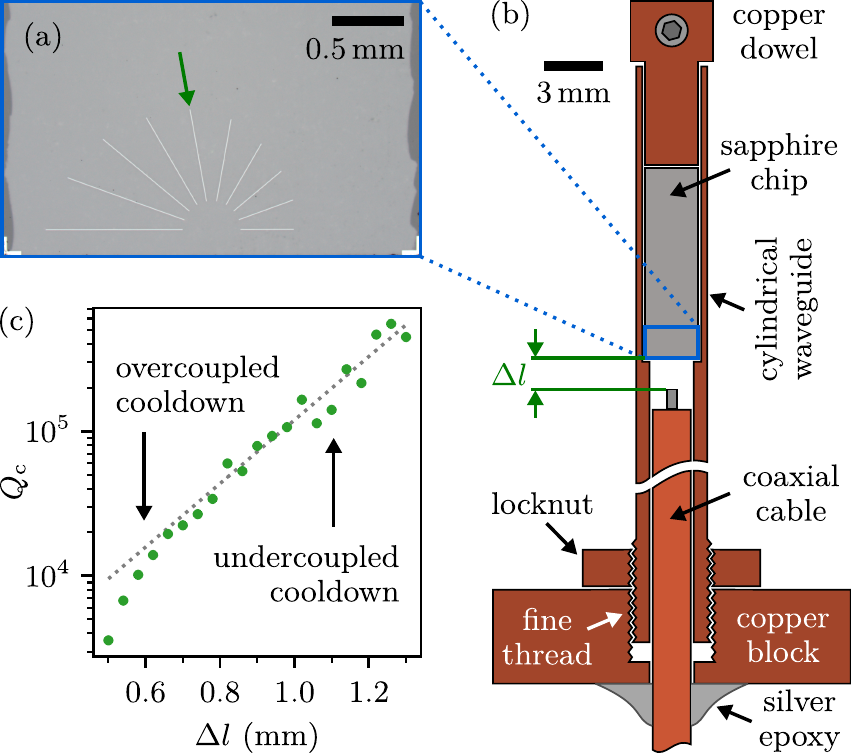}
\caption{\textbf{Chip layout and sample holder.} \textbf{(a)}~Optical image of the sample consisting of 10 stripline resonators in a radial pattern with varying length between \SI{400}{\micro\meter} and \SI{950}{\micro\meter}. \textbf{(b)}~Sample holder schematic, depicting the cross-section through the cylindrical copper waveguide. The design is identical to Refs.~\cite{Borisov2020,Rieger2022}. The chip is fixed inside the waveguide by the copper dowel, which is tightened against the wall of the copper tube. We operate the waveguide below the cut-off frequency ($\sim\SI{60}{\giga\hertz}$) such that the resonators are coupled to the evanescent field of the stripped coaxial cable pin. We can adjust the coupling strength by unscrewing the waveguide with respect to the fixed coaxial cable, which increases the chip-pin distance $\Delta l$. \textbf{(c)}~Finite-element simulation of the coupling quality factor for resonator 5 (green arrow in panel~(a)) depending on the chip-pin distance $\Delta l$. $\Qc$ increases exponentially with $\Delta l$ by approximately one order of magnitude per \SI{0.5}{\milli\meter}, which corresponds to the pitch of the fine thread. The difference in coupling between the two cooldowns (cf.~\tabref{tab:samplePar}) is given by a full turn of the thread.
}
\vspace{-4pt}
\label{fig:supp:samples}
\end{figure}

\begin{table*}[t]
\centering
\caption{\textbf{Sample parameters in the two consecutive cooldowns.} For each resonator, we list the measured resonance frequency $\fr$ and total linewidth $\kappa$ as well as the uncertainty ranges and median values for both $\Qi$ and $\Qc$ calculated with our procedure (cf.~\secref{sec:error_procedure}) for $b^2=\SI{-15}{\deci\bel}$. The chronological order of the experiments was \textit{Undercoupled Cooldown} first, spaced by 3 days at room temperature from the \textit{Overcoupled Cooldown}.}
\label{tab:samplePar}
\vspace{10pt}
\begin{tabular}{c|cccccccc|cccccccc}
\toprule
%\rowcolor{red}
 & & \multicolumn{6}{c}{\cellcolor{lightgreen}Undercoupled Cooldown}  & & & \multicolumn{6}{c}{\cellcolor{lightred}Overcoupled Cooldown} & \\
$\#$ & $f_\text{r}$ & $\frac{\kappa}{(2\pi)}$ & $Q_\text{c, min}$ & $Q_\text{c, mid}$ & $Q_\text{c, max}$ & $Q_\text{i,min}$ & $Q_\text{i,mid}$  & $Q_\text{i,max}$  
 & $f_\text{r}$ & $\frac{\kappa}{2\pi}$ & $Q_\text{c,min}$  & $Q_\text{c,mid}$  & $Q_\text{c,max}$ & $Q_\text{i,min}$ & $Q_\text{i,mid}$  & $Q_\text{i,max}$  \\
 %& & & min & mid & max & min & mid & max & & & min & mid & max & min & mid & max \\
 &$\si{\giga\hertz}$ & $\si{\mega\hertz}$ & $10^3$ & $10^3$ & $10^3$ & $10^3$ & $10^3$ & $10^3$ &$\si{\giga\hertz}$ & $\si{\mega\hertz}$ & $10^3$ & $10^3$ & $10^3$ & $10^3$ & $10^3$ & $10^3$  \\
 \midrule
1 & 5.003  & 0.09  & 179 & 211 & 257 & 70 & 80 & 80 & 4.997  & 0.49  & 12 & 14 & 17 & 30 & 40 & 80 \\ 
2 & 5.644  & 0.09  & 131 & 154 & 187 & 100 & 110 & 130 & 5.641  & 0.63  & 9 & 11 & 13 & 30 & 60 & 790 \\ 
3 & 6.247  & 0.08  & 140 & 164 & 200 & 120 & 130 & 160 & 6.244  & 0.60  & 10 & 12 & 14 & 40 & 80 & - \\ 
4 & 6.848  & 0.06  & 305 & 357 & 429 & 150 & 160 & 170 & 6.846  & 0.31  & 23 & 27 & 33 & 60 & 110 & 430 \\ 
5 & 7.469  & 0.09  & 138 & 162 & 197 & 140 & 170 & 210 & 7.466  & 0.68  & 11 & 12 & 15 & 40 & 110 & - \\ 
6 & 8.273  & 0.16  & 115 & 134 & 161 & 80 & 90 & 100 & 8.269  & 0.87  & 10 & 11 & 14 & 30 & 70 & - \\ 
7 & 9.474  & 0.11  & 250 & 293 & 354 & 120 & 120 & 130 & 9.470  & 0.48  & 20 & 22 & 26 & 80 & 210 & - \\ 
8 & 10.538  & 0.12  & 352 & 414 & 503 & 110 & 120 & 120 & 10.535  & 0.38  & 28 & 33 & 40 & 90 & 160 & 1120 \\ 
9 & 11.310  & 0.09  & 279 & 328 & 397 & 170 & 190 & 210 & 11.307  & 0.52  & 22 & 25 & 30 & 80 & 170 & - \\

\bottomrule
\end{tabular}
\end{table*}

In \figref{fig:supp:samples} we show the sample layout and illustrate how we use the cylindrical waveguide sample holder to change the coupling between the cooldowns (cf.~\tabref{tab:samplePar} and main text discussion). The sample holder design is identical to  Refs.~\cite{Borisov2020,Rieger2022} and couples the samples to the evanescent electric field of a stripped coaxial cable pin. Consequently, we align the grAl resonators in a radial pattern with respect to the pin position (\figrefadd{fig:supp:samples}{a}). Due to its exponential dependence on the chip-pin distance, we can adjust the coupling strength by changing the distance $\Delta l$. We implement this feature by using a (M5x0.5) fine thread  between the copper waveguide tube and the copper block to which the coaxial cable is fixed with silver epoxy glue. Finite-element simulations confirm that $\Qc$ scales exponentially with the chip-pin distance: one order of magnitude per \SI{0.5}{\milli\meter}.

In \tabref{tab:samplePar} we list the extracted parameters for the measurements shown in \figrefadd{fig:examples}{b} at a circulating photon number $\bar{n}\approx 1$. We only use 9 out of the 10 resonators on the chip (cf.~\figrefadd{fig:supp:samples}{a}) because the frequency of the highest resonator is outside the band of the circulator ($\fr>\SI{12}{\giga\hertz}$). Note that, across all resonators, the frequencies have decreased by a few \si{\mega\hertz} from the first cooldown (undercoupled, left hand-side of \tabref{tab:samplePar}) to the second cooldown (overcoupled, right hand-side of \tabref{tab:samplePar}). This observation is consistent with aging of the samples between the cooldowns. We emphasize that both $\Qc$ and $\Qi$ are listed with an uncertainty range and median ($Q_\text{min}$, $Q_\text{mid}$, $Q_\text{max}$), since Fano interference entails a systematic error on the coupling coefficient (cf.~\secref{sec:error_procedure}) and $\Ql=\wr/\kappa$ is measured independently. Note how the uncertainty on $\Qi$ is significantly reduced in the undercoupled cooldown, while for $\Qc$ the uncertainty is lower in the overcoupled case.

\section{Additional Magnetic Field Measurements}
\label{sec:supp:otherfield}
\begin{figure*}[t!]
\centering
\includegraphics[width=\textwidth]{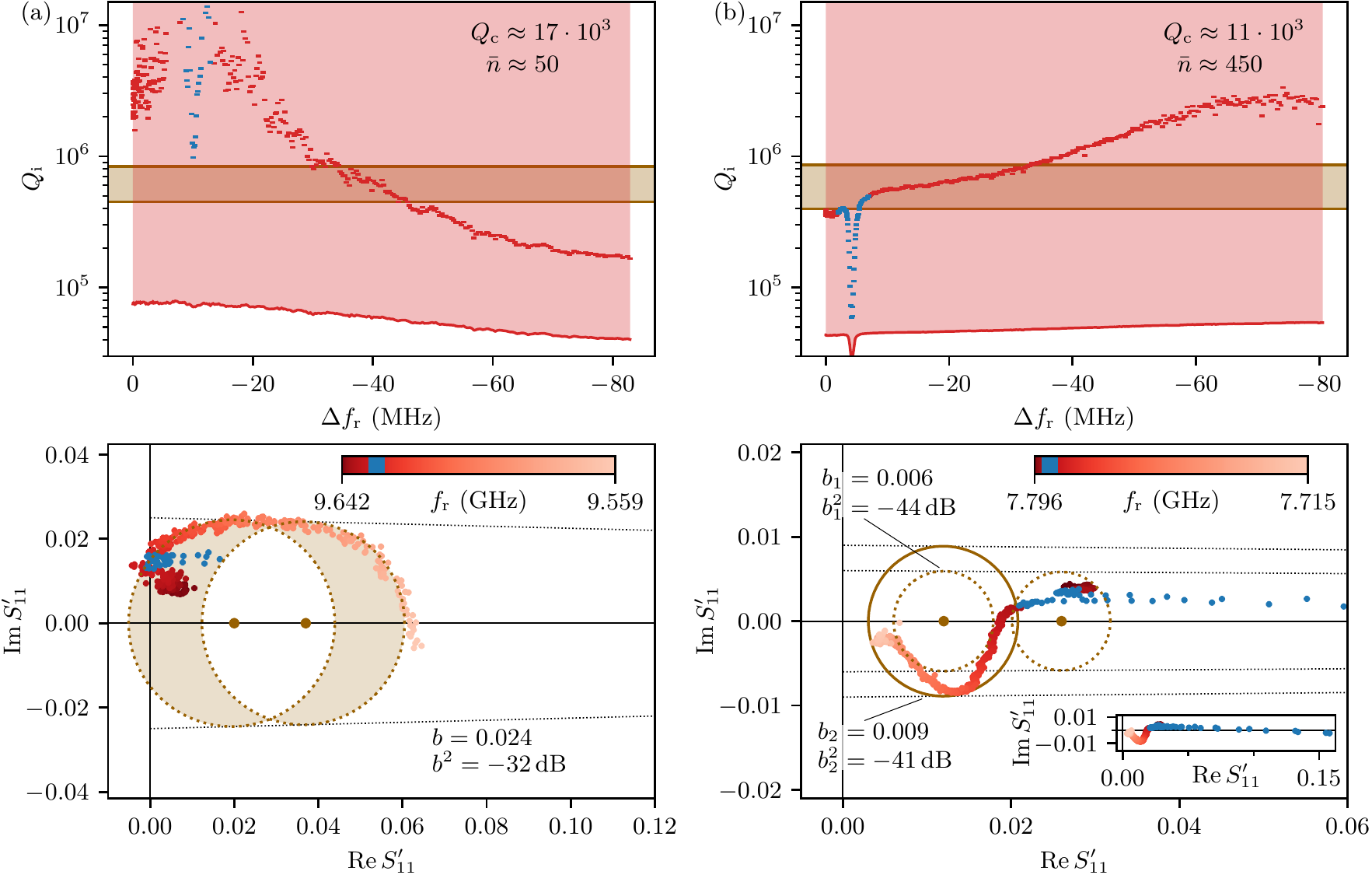}
\caption{\textbf{Resonator frequency sweeps in the overcoupled regime.} Analogous to the experiment presented in \figrefadd{fig:examples}{d}, we use in-plane magnetic field to shift the frequency of grAl resonators over a significant interval compared to the Fano period. In the upper panels, we show the extracted $\Qi$ uncertainty bands and median $Q_\text{i,mid}$. In the bottom panels, we plot the corresponding center points $M'$ of the measured $S'_{11}$ circles in the complex plane. In blue we highlight a \SI{5}{\mega\hertz} span centered on the resonator frequency matching the Zeeman splitting for $g=2$ spins, where we expect microwave losses to spike due to electron spin resonance (ESR). \textbf{(a)} Similar to the main text data (\figrefadd{fig:examples}{d}), $Q_\text{i,mid}$ shows a (deceptive) decrease for the majority of the sweep. Here, only part of the center points follow circular trajectories in the complex plane. We indicate two circles of center points (dashed ocher) at small and large frequency shifts (light red and light orange, respectively), and shade the corresponding range of $\Qi$ in the top panel in ocher. In between the circle sections (orange), the center points follow the straight bounding line for $b=0.024$ (dashed black) towards the off-resonant point $(1,0)$, which corresponds to a decrease of $\Qi$. Note that the fluctuations (dark red) and dip (blue) in $\Qi$ also appear as clusters of points on line trajectories towards~$(1,0)$. \textbf{(b)} In contrast to (a), $Q_\text{i,mid}$ shows a (deceptive) increase by one order of magnitude for the majority of the sweep. In the complex plane (bottom panel) most points follow a clockwise trajectory (red to light orange), which we bound with two circles of center points for different leakage amplitudes $b_1=0.006$ and $b_2=0.009$ (dotted and solid ocher, respectively). We propose an additional $M'(\varphi)$ circle serving as a lower bound for $\Qi$ and covering the initial part (dark red) of the data (with the exception of the ESR dip). The inset shows a zoomed-out version of the data, covering the full extent of the $\Qi$ decrease (blue) due to the ESR with $g=2$ spins. Note that this data corresponds to the measurements shown in Ref.~\cite{Borisov2020}, Fig.\,2 for resonator~B at $\bar{n}=450$.}
\label{fig:supp:otherfield}
\end{figure*}
In \figrefadd{fig:examples}{d} and the main text, we discuss a frequency sweep for which the leakage amplitude $b$ is constant and $\Qi\approx\text{const.}$ holds true for most of the sweep. As a consequence, the center points $M'(f_\mathrm{r})$ trace out a circle in the complex plane (cf.~\figrefadd{fig:fanoTransform}{c}). In \figref{fig:supp:otherfield} we show additional grAl resonator frequency sweeps, in which $b, \Qi = \text{const.}$ do not hold, leading to non-trivial center point trajectories in the complex plane. Similar to \figrefadd{fig:examples}{d}, we sweep the resonator frequency significantly compared to the Fano period by applying a magnetic field up to $\SI{1.2}{\tesla}$ in the plane of the resonator (cf.~\cite{Borisov2020}). In \figrefadd{fig:supp:otherfield}{a}, the elliptic trajectory of the $M'(f_\mathrm{r})$ is consistent with a constant leakage amplitude $b$ and decreasing $\Qi$. Moreover, \figrefadd{fig:supp:otherfield}{b} illustrates the most general scenario of both varying leakage amplitude and varying $\Qi$. We note that in \figrefadd{fig:supp:otherfield}{a,b} the $\Qi$ values at small frequency shifts appear to fluctuate more since they are measured over a considerably longer time span due to the quadratic dependence of resonator frequency shift on magnetic field.

In summary, analyzing the center point trajectory of frequency sweeps is a technique to identify the contribution of Fano interference to deceptive trends in $Q_\text{i,mid}$. However, we emphasize that disentangling changes in $\Qi$ from changes of the Fano amplitude is generally challenging in cases with $b,\Qi \neq \text{const.}$. This limitation of frequency sweeps would be remedied by using a tunable phase-shifter, which allows to vary the phase of the Fano phasor without changing the resonant frequency.

\section{Circle of Center Points}
\label{sec:supp:circleofcenterpoints}
In the following, we derive the equation for the circle of center points $M'(\varphi)$ (cf.~\figrefadd{fig:fanoTransform}{c}). The center point $M$ of $S_{11}$ in absence of Fano effect is located on the x-axis of the complex plane, $M= (x, 0)$. This point is transformed to $M'$ according to \eqref{eq:fanoTransform},
\begin{align*}
    M'=\frac{x + \tilde{b}\e^{\i\varphi}}{1+\tilde{b}\e^{\i\varphi}} \,. %\label{eq:centerpointTransform}
\end{align*}
Isolating the phasor $\tilde{b} \e^{i\varphi}$, taking the absolute value and using $ M' = (x' , y')$ yields
\begin{align}
    &(x'-x)^2 + y'^2 = \left[(1-x')^2 + y'^2\right] \tilde{b}^2\,.
\end{align}
By collecting $x'$ and $y'$ terms and completing the square, we arrive at the equation for the circle of center points,
\begin{align}
    \left(x' - x_c\right)^2 +y'^2 = r_c^2\,,\label{eq:|centerpointTransform|}
\end{align}
with radius and center point on the x axis:
\begin{align}
    r_\text{c} = (1-x)\frac{\tilde{b}}{1-\tilde{b}^2}\label{eq:M_phi_radius}
\end{align}
\begin{align}
    x_\text{c} = \frac{x-\tilde{b}^2}{1-\tilde{b}^2}= 1 - \frac{1-x}{1-\tilde{b}^2}\label{eq:M_phi_center}\,.
\end{align}
Note, that $x_c$ is shifted to the left compared to the original position $x$, i.e. $M$ is not the center of the $M'(\varphi)$ circle. Also note that by identifying $R=1-x$, it is evident that \eqref{eq:M_phi_radius} and \eqref{eq:M_phi_center} describe a homothetic dilation of the $M'(\varphi)$ circle with respect to the off-resonant point. Finally, the angle $\beta$ between the real axis and the lines bounding the $M'(\varphi)$ circles (cf.~\figrefadd{fig:fanoTransform}{d} and \eqref{eq:FanoConeAngle}) is given by
\begin{equation}
    \sin\beta=\frac{r_\text{c}}{1-x_\text{c}}=\tilde{b}
\end{equation}
\end{document}